\documentclass[twocolumn]{revtex4}
\usepackage[dvipdfmx]{graphicx}
\usepackage{amsmath,amsbsy,amssymb}
\usepackage{bm}
\usepackage{mathrsfs}
\usepackage{ulem}
\usepackage{textgreek}

\usepackage{color}

\newcommand{\diff}{\mathrm{d}}
\newcommand{\imag}{\mathrm{Im}\,}
\newcommand{\real}{\mathrm{Re}\,}

\newcommand{\imu}{\mathrm{i}}
\newcommand{\epn}{\mathrm{e}}

\newcommand{\dg}{\dagger}
\newcommand{\la}{\langle}
\newcommand{\ra}{\rangle}
\newcommand{\al}{\alpha}
\newcommand{\sg}{\sigma}

\newcommand{\ep}{\varepsilon}

\begin{document}

\title{
Theory of magnetic skyrmion glass
}

\author{
Shintaro Hoshino${}^{1}$ and Naoto Nagaosa${}^{1,2}$
}

\affiliation{
${}^{1}$RIKEN Center for Emergent Matter Science (CEMS), Wako, Saitama 351-0198, Japan \\
${}^{2}$Department of Applied Physics, The University of Tokyo, Bunkyo, Tokyo 113-8656, Japan
}

\date{\today}

\begin{abstract}
Skyrmions and skyrmion crystal (SkX) discovered in chiral magnets show unique physical 
properties due to their nontrivial topology such as the stability against the annihilation 
and the motion driven by the ultralow current density, which can be advantageous for 
the device applications such as magnetic memories. Especially, the chiral dynamics, i.e., 
the velocity perpendicular to the force acting on a skyrmion, is a key to avoid the impurity 
potential and enhances its mobility. However, the collective pinning of SkX occurs by the 
disorder, which is crucial for its low energy properties. 
Here, we study theoretically the 
low energy dynamics of SkX in the presence of disorder effects in terms of replica field theory, 
and reveal nonreciprocal collective modes and their electromagnetic responses along the 
direction of external magnetic field.
The physical quantities such as the relaxation rate of 
$\mu$SR/NMR 
and 
the pinning frequency
show a dramatic 
change associated with the topological phase transition from the helical state to SkX.
These results provide a firm 
basis to explore the glassy state of SkX.
\end{abstract}

\maketitle

\section{Introduction}

The spin orderings in magnets offer many interesting physics. The typical magnets are the ferromagnets or antiferromagnets with the collinear structures. In these cases, the ground state of the spin system as well as the electronic state are rather simple and well understood. On the other hand, noncolinear spin configurations in magnets are the focus of recent intensive researches, which include the nontrivial ground states such as the helical/spiral magnetic state, and the excited states such as the domain walls, vortices, and magnetic bubbles \cite{nagaosa13,malozemoff_book,chikazumi_book}. The magnetic field or its gradient is a possible way to drive these spin textures \cite{malozemoff_book,chikazumi_book}, while the recent development is their current-driven motion induced by the spin transfer torque in the metallic systems \cite{slonczewski96,berger96,tatara08}. Therefore, the close relation between the electronic conduction and magnetic texture has become an important issue. 
\\ \indent
In particular, the chiral magnets, which have an antisymmetric interaction between spins known as Dzyaloshinskii-Moriya (DM) interaction \cite{dzyaloshinskii58,moriya60}, show a variety of intriguing phenomena caused by nontrivial spin configurations \cite{nagaosa13}.
The most natural one is the helimagnetic state (HMS) since the DM interaction prefers
the winding of spins. Under the external magnetic field $\bm{B}$, three helical spin structures 
with wavevectors perpendicular to $\bm{B}$ are superposed to result in the
triangular crystal of skyrmion, i.e., skyrmion crystal (SkX) \cite{bogdanov89,bogdanov94,roessler06}.
SkX 
has been experimentally identified by using the techniques such as neutron scattering \cite{muhlbauer09}, Lorentz transmission electron microscopy \cite{yu10}, scanning tunneling spectroscopy \cite{heinze11}, and
is now observed in many chiral magnets such as metallic MnSi, (Fe,Co)Si, FeGe \cite{jonietz10,heinze11,yu11,schulz12}.
These materials have recently been intensively investigated because of their topological properties:
noncolinear spin textures of SkX have nontrivial winding number.
While their basic properties have been well understood for pure systems, 
in real materials there are always disorders from e.g. 
crystal defects or impurities. This is clear from the finite threshold critical current 
density $j_{\rm c}$ to derive the motion by spin transfer torque effect, i.e., the impurity 
pinning \cite{jonietz10}. It is expected that the disorder modifies the low energy dynamics of the
SkX, which can be studied by X-ray diffraction, $\mu$SR resonance, NMR,
optical conductivity, and ac magnetic susceptibility.
The magnetic SkX appears in the presence of spin-orbit coupling and both time-reversal 
and spatial inversion symmetry breaking, and hence they are closely related to 
multiferroic properties. 
Especially, the insulating multiferroic skyrmionic system Cu$_2$OSeO$_3$ has been found \cite{seki12,ozerov14}.
Note also that the glassy behavior in multiferroics has been recently studied in a magnetoelectric insulator 
BaCo$_6$Ti$_6$O$_{19}$ \cite{tonomoto16} where the spin-orbit interaction plays an important role as in chiral magnets.

Pinning phenomenon has been intensively studied for the charge density wave (CDW)
and spin density wave (SDW), vortex lattice in type-II superconductors, and Wigner crystal \cite{fukuyama78,lee79,larkin70,giamarchi94,giamarchi95,giamarchi96,chitra98,chitra01}.
Because of the broken translational symmetry, the corresponding Goldstone modes,
i.e., phasons, govern the low energy dynamics. Phasons can be regarded as the
acoustic phonon of the 
charge, spin, and vortex crystals, respectively. 
There are two types of pinning, i.e., strong pinning and weak pinning.
The former occurs when the impurity pinning strength is very strong and 
the phason is pinned at each impurity, while the weak impurities collectively 
pin the phason in the latter case \cite{fukuyama78,lee79}. In the weak pinning case, there 
appears a typical size of the domain $\xi_{\rm p}$ called pinning length over which the 
phason varies of the order of $\pi$. 
The resultant glassy state is characterized by a power-law dependence in spatial correlation function and magnetic form factors \cite{giamarchi94,giamarchi95}.
This has been known as a Bragg glass state and has been discussed in disordered systems such as the superconducting vortex lattice \cite{klein01}.
Also the typical energy scale, i.e., 
pinning frequency $\omega_{\rm p}$, appears,
and it creates a gap in phason dispersion.

\begin{figure}[t]
\begin{center}
\includegraphics[width=60mm]{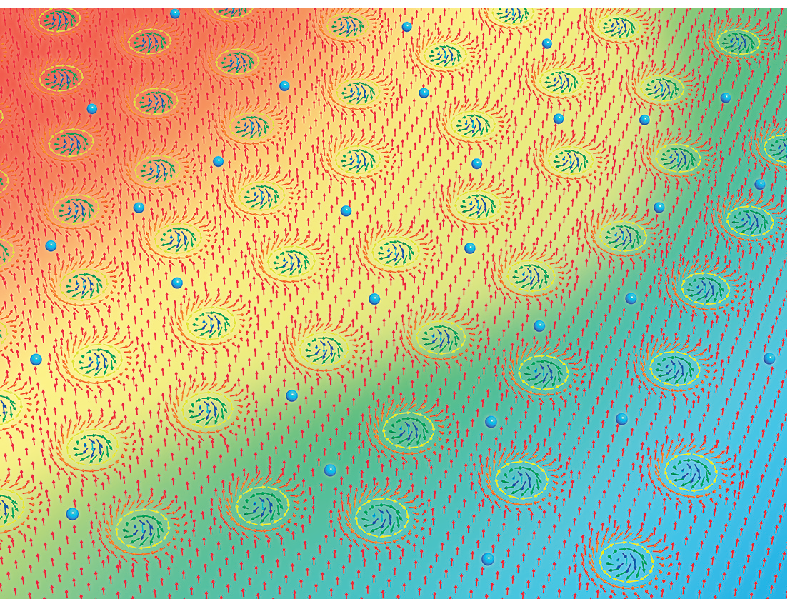}\vspace{2mm}
\includegraphics[width=60mm]{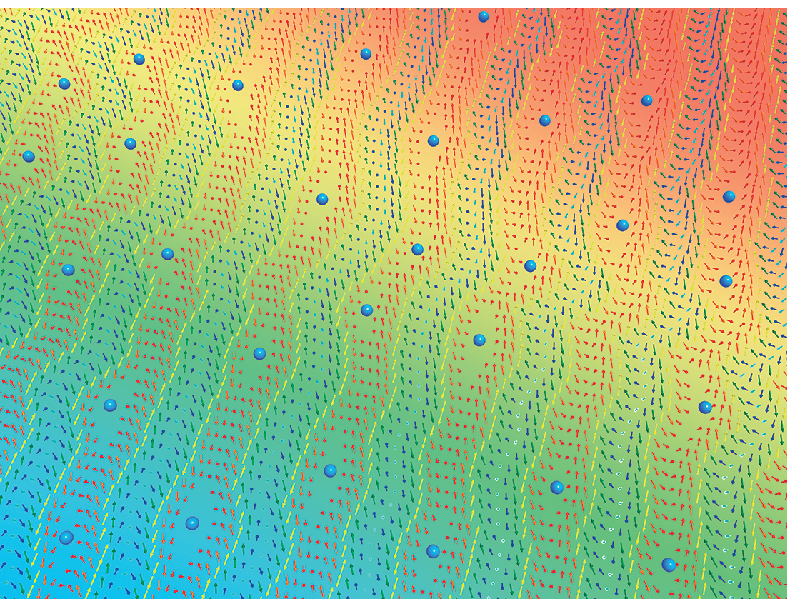}
\caption{
Illustrations for (Top) disordered magnetic SkX and (Bottom) disordered HMS in two dimensions.
Color gradations indicate the spatially varying phase, whose characteristic length scale is $\xi_{\rm p}$ defined in Eq.~\eqref{eq:pin_length}.
}
\label{fig:illust}
\end{center}
\end{figure}

Although these accumulated knowledge is helpful to understand the 
disordered SkX, there are several new features distinct from CDW, SDW,
vortex lattice, and Wigner crystal. (i) The gyrodynamics due to the topological skyrmion 
number relates the $x$- and $y$-components of the SkX displacement 
as the
canonical conjugate pair \cite{zang11}, and hence its dispersion becomes $\omega_{\bm q}
\propto \bm{q}^2$ in sharp contrast to the cases of
CDW and SDW. Although this feature is common with the vortex lattice and Wigner crystal,
the phason is overdamped in the vortex lattice and not well defined, while
it is underdamped in SkX \cite{zang11}. 
Wigner crystal is more similar to SkX, but the magnetic 
properties are unique to SkX.
 (ii) There are nonreciprocal nature of dynamics and electromagnetic responses,
i.e., the difference between parallel and anti-parallel to the external magnetic 
field $\bm{B}$, which is closely related to the chirality of the system. 
This can be detected by e.g. the directional propagation of 
magnetic collective modes as well as the directional dichroism analogous to that
observed in insulating multiferroics \cite{kezsmarki11}.  
(iii) As the external magnetic field $\bm{B}$ increases,
there occurs the first-order topological phase transition from HMS to SkX state, which offers a unique opportunity to compare various physical 
properties in these two cases as shown in Fig.~\ref{fig:illust}. 
Namely, HMS (topologically trivial) is similar to the
conventional SDW, while the SkX is characterized by the nontrivial topology.
Already the orders of magnitude difference in the threshold current density 
$j_c$ has been reported experimentally \cite{jonietz10, schulz12}, and theoretically analyzed \cite{zang11}.
(iv) It is also possible to compare the two-dimensional thin film sample and
three-dimensional single crystal sample of the same material. The pinning 
properties depend strongly on the dimensionality, and this comparison provides 
useful information.

As for the impurity effects on SkX, there have been previous works that discuss the pinning-depinning transition and the movement of skyrmions \cite{reichhardt15,lin12,hanneken16,raposo17,liu13,reichhardt17}.
These works study the non-equilibrium dynamics of skyrmions 
in terms of numerical simulation.
On the other hand, the glassy nature of SkX below the critical current density has still remained to be explored both theoretically and experimentally.
In this paper, we study 
the disorder effect on SkX and HMS
in terms of replica field theory originally developed for spin glass
and later applied to CDW, SDW, vortex lattice, and Wigner crystal \cite{mezard_book,mezard91,giamarchi94,giamarchi95,giamarchi96,chitra98,chitra01}.
The present theory provides predictions on the
domain size $\xi_{\rm p}$, threshold current density $j_{\rm c}$,
pinning frequency $\omega_{\rm p}$, nonreciprocal dispersion of phasons, ac conductivity 
$\sigma_{xx}(\omega)$ and $\sigma_{xy}(\omega)$, ac magnetic susceptibility
$\chi_{\mu \nu}(\bm{q}, \omega)$, and relaxation time $T_1$ in $\mu$SR and NMR, 
in terms of the strength of the impurities and their density.
The comparison between two and three dimensions, together with the
SkX state and HMS are summarized in Tables \ref{tab:expr}, 
\ref{tab:expr2} and \ref{tab:expr3}.

\section{Effective action and Green function for disordered skyrmion crystals}

We begin with the Hamiltonian for
the two ($d=2$) and three ($d=3$)  
dimensional chiral ferromagnets (FM) with Dzyaloshinskii-Moriya (DM) interaction 
in the presence of pinning potentials:
$\mathscr H = \mathscr H_{\rm DMFM} + \mathscr H_{\rm pin}$.
The ferromagnetic part $\mathscr H_{\rm DMFM}$ contains the symmetric Heisenberg exchange $- J \bm S_i \cdot \bm S_j$ between the nearest neighbor spins $\bm S_i$ and $\bm S_j$, and antisymmetric DM interaction $\bm D \cdot (\bm S_i \times \bm S_j)$ with $\bm D$ pointing to the direction along the bond \cite{dzyaloshinskii58,moriya60}.
This Hamiltonian describes chiral magnets with cubic symmetry such as MnSi.
We also consider the external magnetic field $\bm B$.
Assuming that the characteristic length scale for a magnetic texture is much longer than the lattice constant, we obtain the following Hamiltonian \cite{bak80}
\begin{align}
\hspace{-3mm}
\mathscr{H}_{\rm DMFM} &= \int \frac{\diff \bm r}{a^d} \left[
\frac{Ja^2}{2} (\bm \nabla \bm S)^2 + Da\bm S\cdot (\bm \nabla \times \bm S)
-g_s\mu_{\rm B}B S^z
\right]
, \label{eq:ham_fm}
\\
\mathscr{H}_{\rm pin} &= \int \frac{\diff \bm r}{\ell^d} \, V(\bm r) (S^z)^2
, \label{eq:ham_pin}
\end{align}
where $D=|\bm D|$, $g_s\simeq 2$ is the $g$-factor and $\mu_{\rm B}$ is the Bohr magneton.
The magnetic field is applied along the $z$ axis.
The impurity or disorder effect enters through the pinning potential $V$.
The length $a$ is a lattice constant and $\ell$ ($>a$) is an averaged distance between impurities.
In real materials, the impurities or defects do not break the time-reversal symmetry, and then the simplest form of the disorder potential is a form of random magnetic anisotropy which is bilinear in $\bm S$. 
For simplicity, we take only the $z$-component as in Eq.~\eqref{eq:ham_pin}, which pins each skyrmion at its center.
We assume that the impurity potential $V$ has a Gaussian distribution and satisfies
$\overline{V(\bm r)V(\bm r')} = V_{\rm imp}^2 \ell^d \delta(\bm r-\bm r')$
where overline means the impurity average.
We note the relation $V_{\rm imp}^2\simeq \overline{V(\bm r)^2}$.
We expect $V_{\rm imp}\sim J$, since impurity effects considered in this paper may enter through the lattice defect which modifies interactions.
In the following, we concentrate on SkX.
The results for HMS are given in Appendix \ref{sec:appendix1}, and we will compare the results between SkX and HMS in Sec.~\ref{sec:discussion}.

The spin moment for skyrmion spin texture 
is characterized by a 
hexagonal pattern in the magnetic structure factor \cite{muhlbauer09},
and is given by \cite{petrova11,tatara14}
\begin{align}
\bm S &= 
S\sum_{i=1}^3 [ \eta_i \hat {\bm Q_i} + \sqrt{1-\eta_i^2} \bm n_i]
, \label{eq:triple-helix}
\\
\bm n_i &= \hat {\bm z} \cos(\bm Q_i \cdot \bm r + \phi_i)
+ \hat {\bm Q}_i\times \hat {\bm z} \sin (\bm Q_i \cdot \bm r + \phi_i)
,
\end{align}
where the three helical spin structures are superposed.
The wave vectors are defined by 
$\hat {\bm Q}_1 = (-\sqrt 3\hat{\bm x}-\hat{\bm y})/2$, 
$\hat {\bm Q}_2 = ( \sqrt 3\hat{\bm x}-\hat{\bm y})/2$, and
$\hat {\bm Q}_3 = \hat{\bm y}$.
The magnitude $Q\equiv|\bm Q_i|=D/Ja$ characterizes the size of magnetic skyrmions.
Here we have introduced the phason field $\phi_i$, which describes the low-energy dynamics of SkX, and the other field $\eta_i$ is its canonical conjugate variable.
The field variable $\phi_i$ represents a modulation of helix along $\bm Q_i$ where the spins are directed inside the plane perpendicular to $\bm Q_i$, and $\eta_i$ corresponds to a tilt of the spin moment from this plane \cite{petrova11,tatara14}.

The elementary excitation at low energy and at long wavelength is described by two-component phason fields defined by
$\phi_x = (-\phi_1 +\phi_2 )/\sqrt 2$ and $\phi_y = (-\phi_1-\phi_2+2\phi_3)/\sqrt 6$, which are regarded as local displacement fields by multiplying $Q^{-1}$.
The similar linear combinations also apply to $\eta_i$, which defines $\eta_x$ and $\eta_y$.
There is also the symmetric (breathing) part 
$\phi_s = (\phi_1+\phi_2+\phi_3)/\sqrt 3$, but
its dynamics is absent in the low-energy theory of phasons and is not important
\cite{tatara14}.
Furthermore, the DM interaction along $z$ direction, which is a source of nonreciprocity discussed in this paper, does not have influence on this symmetric part of phason.
For these reasons we here neglect it.

Assuming that the spatial modulation of $\phi_i$ and the magnitude of $\eta_i$ are small, we can write
the low-energy effective Hamiltonian 
as 
\begin{align}
&\hspace{-3mm}
\mathscr{H}_{\rm DMFM} \simeq
\int \frac{\diff \bm r}{a^d} \sum_{\al\beta} \left[
\frac{JS^2 a^2}{2} \delta_{\al\beta}(\bm \nabla \phi_{\al})^2 + 
\frac 1 2 \eta_{\al} g_{\beta\al}^{-1} (-\imu\bm \nabla) \eta_{\beta}
\right]
, \label{eq:ham_eff}
\\
&\hat g^{-1}(\bm q) =
JS^2 a^2Q^2 \hat 1
+ 3 \imu DS^2 a q_z \hat \epsilon
,\label{eq:g0_func}
\end{align}
where 
the suffices $\al,\beta$ indicate $x$ and $y$.
We have introduced the antisymmetric tensor $\epsilon_{xx}=\epsilon_{yy}=0$, $\epsilon_{xy}=-\epsilon_{yx}=1$.
Putting the Berry phase term with imaginary time representation \cite{zang11,tatara14,zhang16}
\begin{align}
\mathscr{S}_{\rm B} &= \hbar S \int \frac{\diff \bm r}{a^d} \sum_{\al\beta} \int _0^{\hbar/k_{\rm B}T}
\hspace{-3mm} \diff \tau \,
 \left[
\frac{9\imu}{8} \phi_{\al} \epsilon_{\al\beta}
\dot \phi_{\beta}
- \frac{3\imu }{2} \eta_{\al}\delta_{\al\beta} 
\dot \phi_{\beta}
\right]
\label{eq:Berry_phase}
\end{align}
together, we obtain the effective action in the low-energy limit as
$\mathscr{S} = \mathscr{S}_{\rm B} + \int \diff \tau 
\mathscr{H}_{\rm DMFM}
$ for clean systems.
Here $T$ is a temperature.
The first term in Eq.~\eqref{eq:Berry_phase} is specific to SkX \cite{zang11}.
The
 topological property characterized by the number 
$N_{\rm SkX} = (4\pi)^{-1} \int \diff A \, \bm n \cdot (\partial_x \bm n \times \partial_y \bm n)$ is responsible for this term, where the integral $\int \diff A$ is performed over the two-dimensional area and $\bm n=\bm S/S$.
With this low-energy effective action,
the partition function is given by
$Z = \int 
\mathscr{D} \{\phi_{\al}\} \mathscr{D}\{\eta_{\al}\} \epn^{-\mathscr{S}/\hbar}
$.
The functional integral with respect to $\eta_{\al}$ can be performed without approximations, and then only the phason modes need to be considered.

By combining the pinning potential given in Eq.~\eqref{eq:ham_pin} with the above model, we can obtain the effective action that determines the low-energy dynamics in the presence of disorder effects.
Using the replica field theory combined with a Gaussian variational approximation \cite{mezard91,giamarchi94,chitra98}, we obtain the response functions, whose derivations are summarized in Appendix \ref{sec:appendixNEW}.
Thus the phason Green function $\mathscr G_{\al\beta} (\bm q,\omega)= \overline{\la \phi_{\al}(\bm q,\omega) \phi_{\beta} (-\bm q,-\omega) \ra}/a^d\hbar$
is given by
\begin{align}
\hat{ \mathscr{G}}^{-1} (\bm q,\omega)
&= \left[
JS^2a^2 \bm q^2 - \frac{9J\hbar^2}{4D^2}\omega^2 + \Sigma_1 - \imu \Sigma_2\hbar \omega
\right] \hat 1
\nonumber \\
&\ \ + \left[
\frac{9 \imu S\hbar}{4}\omega
+ \frac{27 \imu J^2 \hbar^2 a}{4D^3} \omega^2 q_z
\right] \hat \epsilon
, \label{eq:green}
\end{align}
for $\omega>0$, where we have introduced the static self-energy $\Sigma_1$ and the damping part $\imu \Sigma_2 \hbar \omega$ valid for 
$\omega \rightarrow 0$.
The $q_z$-linear term in Eq.~\eqref{eq:green} (also in Eq.~\eqref{eq:g0_func}) appears only for $d=3$ and vanishes for $d=2$.
It represents the 
mirror
symmetry breaking along $z$ (magnetic field) direction.
Note that this term is characteristic for SkX where three helices are superposed, and is not found in the single helix state.
With use of this Green function, we can also explicitly write down the Green function $\mathscr F_{\al\beta} = \overline{\la \eta_{\al} \eta_{\beta} \ra} /a^d\hbar$ for the $\eta$-field as
\begin{align}
\hat {\mathscr F}(\bm q,\omega)
&= \hat g(\bm q) + \left( \frac{3S\hbar\omega \hat g(\bm q)}{2} \right)^2 \hat {\mathscr G}(\bm q,\omega)
,\label{eq:Green_beta}
\end{align}
where $\hat g(\bm q)$ corresponds to a static susceptibility.
Equation~\eqref{eq:Green_beta} can be derived if one starts from the action before performing the integration with respect to $\eta$-field.
The self-energy coefficients are
\begin{align}
\Sigma_1 &= \left( \frac{3 V_{\rm imp}^2S a^3}{4\pi J^{3/2} \ell^3} \right)^2
, \label{eq:Sigma1_3d}
\\
\Sigma_2 &= \sqrt{\frac{9J\Sigma_1}{D^2} + \left(\frac{9S}{4}\right)^2}
\simeq \frac{9S}{4}
, \label{eq:Sigma2_3d}
\\
\Sigma_1^{\rm (2D)} &= \frac{3V_{\rm imp}^2S^2 a^2}{2\pi J \ell^2}
, \label{eq:Sigma1_2d}
\\
\Sigma_2^{\rm (2D)} &= \sqrt{\frac{9J\Sigma_1}{2D^2} + \left(\frac{9S}{4}\right)^2}
\simeq \frac{9S}{4}
. \label{eq:Sigma2_2d}
\end{align}
We have neglected the high-order $\omega^2q_z$ term in evaluating the self-energy.
In the final expression for $\Sigma_2$, we have kept the leading order term by taking the weak disorder limit as $\Sigma_1\rightarrow 0$.

The form of the damping coefficient $\Sigma_2$ in Eqs.~\eqref{eq:Sigma2_3d} and \eqref{eq:Sigma2_2d} originates from the first term in Eq. \eqref{eq:Berry_phase} related to the topological property of SkX.
The value $\Sigma_2 \sim S$ is much enhanced compared to the case without this term, i.e. the case of HMS (see Appendix~\ref{sec:appendix1}), for weak impurities.
This is due to the dramatic change in the energy dispersion for phasons in the presence of the topological term (see Sec.~\ref{sec:dispersion} and Fig.~\ref{fig:dispersion}), which creates a large number of low-energy states that are involved in the damping process.
We note that the damping becomes zero if we take the impurity potential $V_{\rm imp}$ as zero, since this $\omega$-linear form of the self-energy is derived for frequencies much smaller than the characteristic pinning frequency $\omega_{\rm p}$ which goes to zero when $V_{\rm imp}\rightarrow 0$ as discussed in the next.

The static part of self-energy introduces new energy and length scales.
To see these quantities, we first consider the Euler-Lagrange equation of motion
$\sum_{\beta} \mathscr G^{-1}_{\al\beta} (\imu \bm \nabla, -\imu \partial_t) \phi_{\beta} (\bm r,t) = 0$.
For the low-lying excitation mode without spatial modulation,
the equation of motion has the form of the damped oscillator 
$\ddot \phi_{\al} + 2\zeta \omega_{\rm p} \dot \phi_{\al} + \omega_{\rm p}^2 \phi_{\al} =0$ 
where $\omega_{\rm p}$ and $\zeta$ ($>0$) are pinning frequency and dimensionless damping ratio.
The specific forms of these constants are
\begin{align}
\hbar \omega_{\rm p} = \frac{2\sqrt 2 \Sigma_1}{9S}
, \ \ 
\zeta = \frac{1}{\sqrt 2}
.
\end{align}
The resonance frequency is $\sqrt{1-\zeta^2}\omega_{\rm p}$ which can be defined for $\zeta<1$, and the damping rate is given by $\zeta\omega_{\rm p}$.
Namely, the phasons for SkX are located in the underdamped regime near the critically damped case at $\zeta=1$.
These expressions are valid for both $d=3$ and $d=2$, and the explicit parameter dependences of pinning frequencies are summarized in Tab.~\ref{tab:expr}.
The pinning frequency for $d=3$ ($\omega_{\rm p}\propto V_{\rm imp}^4\ell^{-6}$) is strongly dependent on $V_{\rm imp}$ and $\ell$, and is smaller than the one in the two-dimensional case ($\omega_{\rm p}\propto V_{\rm imp}^2\ell^{-2}$) for weak impurity potentials.
Intuitively this behavior can be understood as follows: in two dimensions the skyrmion can be pinned by a point-like pinning center. On the other hand, in three dimensions, we have the skyrmion strings each of which is pinned by impurities, but the segment between pinned points can be modulated.
Thus the pinning is effectively much weaker in three dimensions.

The static part $\Sigma_1$ also defines a new characteristic length scale called pinning length $\xi_{\rm p}$ given by
\begin{align}
\xi_{\rm p} &= a \sqrt{\frac{JS^2}{\Sigma_1}}
\sim a \left( \frac{J \ell^{d/2}}{V_{\rm imp}a^{d/2}} \right)^{\frac{2}{4-d}}
, \label{eq:pin_length}
\end{align}
which is also listed in Tab.~\ref{tab:expr}.
We can show that this length scale represents the collective pinning by weak impurities \cite{fukuyama78,lee79,giamarchi96}.
To demonstrate this in the context of magnets, we first introduce the energy density from the spatial modulation with the length scale $\xi$ caused by impurities:
\begin{align}
E(\xi) &= \frac{JS^2}{\xi^2a^{d-2}} - \frac{\sqrt{(V_{\rm imp}S^2)^2 n_{\rm imp}\xi^d }}{\xi^d }
. \label{eq:energy_pin}
\end{align}
Here $n_{\rm imp}\sim \ell^{-d }$ is the impurity density.
The first and second terms are energy loss of the ferromagnetic interaction and energy gain due to the impurity potential, respectively.
One can see that the pinning length $\xi\sim\xi_{\rm p}$ minimizes this energy density.
Thus the present Gaussian variational approximation used in the replica field theory is valid for weak-pinning regime with $\xi_{\rm p} \gg \ell$.
From Tab.~\ref{tab:expr}, we have the ratio $\xi_{\rm p}/\ell\sim (V_{\rm imp}/J)^2(a/\ell)^2$ in $d=3$, which indicates that the system is located in the weak pinning regime for $a\ll \ell$ and $V_{\rm imp}\sim J$.
In $d=2$, we have the relation $\xi_{\rm p}/\ell \sim V_{\rm imp}/J$ and there is no $a/\ell$ factor.
Thus the pinning length for $d=3$ is much longer than that for $d=2$, corresponding to the smaller pinning frequency in $d=3$ than 
in $d=2$.

At the end of this section, we comment on a nature of the Bragg glass.
While the static properties are basically similar to the ones in the previous works for vortex lattice \cite{korshunov93,giamarchi95}, let us take a look at this behavior based on our model.
The mean squared phason field variable, meaning the roughness, are calculated as $B(\bm r)\equiv \overline{\la [\phi_{\al} (\bm r) - \phi_{\al} (\bm 0)]^2\ra} \simeq (4-d) \ln (r/\xi_{\rm p})$ at large distance with $r\gg \xi_{\rm p}$.
This logarithmic growth is characteristic for the Bragg glass.
With this quantity, we can obtain the impurity-averaged spin correlation function as $\overline{\la \bm S(\bm r)\cdot \bm S(\bm 0) \ra} \simeq 3S^2 (r/\xi_{\rm p})^{-(4-d)/3}$ which shows the power-law decay.
Because of this slowly decaying property, the Bragg glass state is referred to as a quasiordered state \cite{giamarchi95}.
Correspondingly, the Fourier transformed quantity also shows a power-law behavior, which can be measured by elastic neutron scattering measurement as in the vortex lattice \cite{klein01}.
For chiral magnets, the pinned crystal states have been identified experimentally \cite{muhlbauer09,schulz12} and theoretically \cite{reichhardt15} at small pinning potential, which is expected to be the skyrmion Bragg glass discussed in this paper.
In contrast, for a perfect crystal $B(\bm r)\simeq {\rm const.}$ is satisfied and the Fourier transformed magnetic form factor becomes a delta function in $\bm q$-space.
With stronger disorder effects, on the other hand, we can have a power-law increase in the function $B(\bm r)$, which results in an exponential decay of the spin correlation functions.
These two cases are clearly distinguished from the Bragg glass state.

The existence of Bragg glass has been debated over the years.
While there have been supports for the existence after the proposal of this state in three dimensions \cite{gingras96, fisher97, klein01}, the renormalization group theory suggested the absence of the Bragg glass even in the three dimensions: the perturbative analysis for the XY model shows the lower-critical dimension $d_{\rm lc}\simeq 3.9$ \cite{doussal06,tissier06-2}, which is far beyond $d=3$.
There is a counter argument that this conclusion relies on the perturbative analysis around the critical point and the higher-order perturbations will modify the critical dimension into $d_{\rm lc}=2$ \cite{wiese06}. 
On the other hand, the non-perturbative scaling analysis has also been performed to give $d_{\rm lc}\simeq 3.8$ \cite{tissier06,tissier06-2}.
Here, the truncated version of functional renormalization group equations has been used, which however may differ from results obtained for the non-truncated version \cite{wiese06}.
Recently, the existence of Bragg glass has been further supported by theoretical and experimental investigations for density-wave glasses \cite{okamoto15,okamoto15-2,mross15}.

In two dimensions, on the other hand, it is naively expected that the generation of dislocations, or vortices in phason field variables, can occur under the presence of disorders and can change the picture of quasiordered state 
\cite{gingras96,carpentier98,zeng99}.
However, the characteristic length scale for dislocations is shown to be much longer than the pinning length, and there can be a wide region where the Bragg glass behavior is observed \cite{giamarchi95, carpentier98, doussal00}.
Indeed, the pinned Wigner crystal state has been interpreted as a quasi-Bragg glass state in two dimensions \cite{chitra01}.
In $d=3$, the dislocation loop energetically costs much more than the $d=2$ case, and the Bragg glass state 
can be more
robustly present \cite{giamarchi95}.

\section{Nonreciprocity in dispersion relation} \label{sec:dispersion}

\begin{figure}[t]
\begin{center}
\includegraphics[width=90mm]{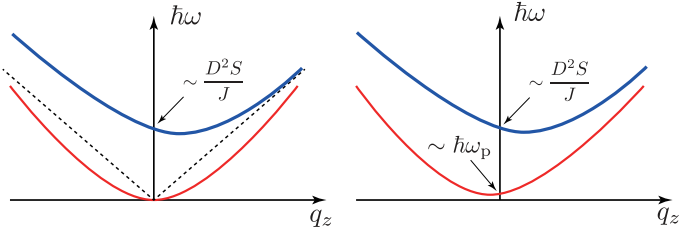}
\caption{
Schematic pictures of dispersion relations along $q_z$ direction for the three dimensional SkX in (Left) clean case and (Right) dirty case.
The dispersion relation in the Right panel is determined by the peak positions of the response functions.
}
\label{fig:dispersion}
\end{center}
\end{figure}

Here we discuss the dispersion relation of phasons, which is determined by the poles of the Green functions in Eq.~\eqref{eq:green}.
Let us first consider the clean case without impurities.
For the three dimensional SkX, we have two kinds of excitation modes and can write the dispersions at small wave vectors as
\begin{align}
\hbar \omega_{\rm low}(\bm q) &\simeq \frac{4 Ja^2S}{9} \bm q^2
- \frac{16 J^3a^4S}{81D^2} \bm q^4
- \frac{16 J^4a^5 S}{27 D^3} \bm q^4q_z
+O(q^6)
, \label{eq:omega1}
\\
\hbar \omega_{\rm high}(\bm q) &\simeq \frac{D^2S}{J} 
+ 3DaS q_z
+ \frac{4 Ja^2S}{9} \bm q^2 + 9J a^2S q_z^2
+O(q^3)
. \label{eq:omega2}
\end{align}
The quadratic form in the low-energy branch at small $\bm q$ is characteristic for SkX \cite{zang11}, and the schematic illustration for the dispersion relations are shown in the left panel of Fig.~\ref{fig:dispersion}.
This behavior originates from the Berry phase term connecting the field variables $\phi_x$ and $\phi_y$ in Eq.~\eqref{eq:Berry_phase}, with which a rotational motion is generated \cite{zang11}.

In Eqs.~\eqref{eq:omega1} and \eqref{eq:omega2},
 we keep the lowest-order terms with respect to $q_z$, which represents the nonreciprocity along magnetic field ($z$) direction characteristic for SkX in $d=3$.
For the lower energy branch $\omega_{\rm low}$, the $q_z$ term appears in the form $\bm q^4 q_z$.
This is because the presence of $q_z$-linear term makes the energy negative to cause instability, and is not allowed.
For the higher $\omega_{\rm high}$ branch, on the other hand, the $q_z$-linear term can appear.
The minimum of the frequency $\omega_{\rm high}(\bm q)$ is lowered by this $q_z$-linear term, and the energy shift is estimated as
\begin{align}
\hbar \varDelta \omega_{\rm high} &\sim \frac{D^2S}{J}
\end{align}
which is similar to the ferromagnetic state (see Appendix  \ref{sec:appendix3}).
If this bottom of the dispersion relation touches the zero energy by e.g. tuning DM interaction along $z$-direction, the instability toward modulation along $z$ direction will occur.
For two dimensions, we get the dispersion relations by disregarding the $q_z$-dependent terms, and nonreciprocal nature does not arise in the $xy$-plane \cite{tatara14}.
The absence of nonreciprocal dispersion applies also to HMS (see Appendix \ref{sec:appendix1}).
Thus the $q_z$ terms in the dispersion relation in three dimensional SkX shows a sharp contrast with SkX in $d=2$ and HMS in $d=2,3$.

In the presence of impurities,
the lower excitation spectrum has a peak at finite frequency, and the dispersion relation $\omega_{\rm low}'(\bm q)$ is defined as
\begin{align}
\hbar \omega_{\rm low}' (\bm q) &\simeq 
\hbar\omega_{\rm p}
- \frac{4\sqrt 2 J^2 a \Sigma_1^2}{27 D^3 S^3} q_z
+\frac{4 Ja^2S}{9} \bm q^2 
+O(q^3)
, \label{eq:omega1p}
\end{align}
where we have neglected the attenuation to visualize the dispersion relation.
The $q_z$-linear term can now be present since the energy gap is generated from impurity pinning, and the excitation energy remains finite even with this term.
This is illustrated in the right panel of Fig.~\ref{fig:dispersion}.
The shift of minimum position in the $\omega_{\rm low}'$ branch due to the $q_z$-linear term is estimated as
\begin{align}
\hbar \varDelta \omega_{\rm low}' \sim 
\frac{J^3\Sigma_1^4}{D^6S^7}
\end{align}
which is very small compared to $\hbar \omega_{\rm p}$ in the weak disorder case.
For $\omega_{\rm high}$, the effect of impurities does not enter in the leading order.

Physical intuition about these excitation modes can be obtained by
analyzing the equations of motion in the uniform limit with $\bm q=\bm 0$.
These equations are easily solved, and
one can see that the above two modes $\omega_{\rm low}'$ and $\omega_{\rm high}$ correspond to the clockwise and counter clockwise motions with the frequencies $\omega_{\rm low}'=\omega_{\rm p}$ and $\omega_{\rm high}=\frac{D^2S}{\hbar J}$.
These clockwise and counter clockwise natures are also reflected in the signs of the $q_z$-linear terms as in Eqs.~\eqref{eq:omega2} and \eqref{eq:omega1p}.

\section{Response functions}

\subsection{Local magnetic response}

\begin{figure}[t]
\begin{center}
\includegraphics[width=85mm]{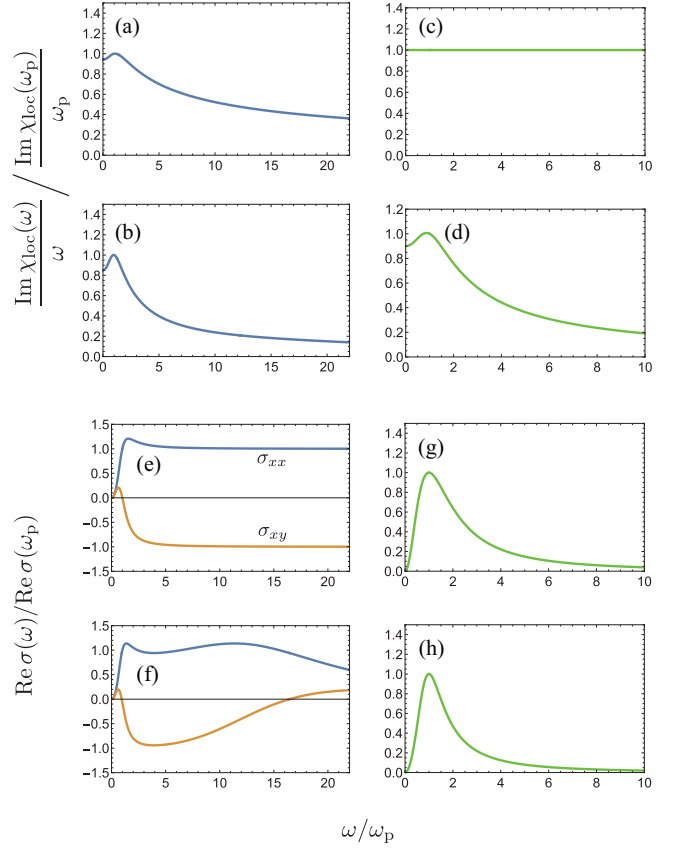}
\caption{
Frequency 
dependences of local magnetic response functions for (a) SkX in $d=3$, (b) SkX in $d=2$, (c) HMS in $d=3$, and (d) HMS in $d=2$. 
The electrical conductivities are also shown for (e) SkX in $d=3$, (f) SkX in $d=2$, (g) HMS in $d=3$, and (h) HMS in $d=2$.
The typical parameters for MnSi are used for the 
plots (see Sec.~\ref{sec:discussion}).
}
\label{fig:plots}
\end{center}
\end{figure}

Let us
 consider the magnetic response of SkX.
We first define the real-space dynamical magnetic susceptibility by
$\chi^{\mu\nu}(\bm r,\bm r',t) = 
\overline{
\la S^\mu (\bm r,t) S^\nu (\bm r', 0) \ra 
}
$.
We can rewrite it in terms of dynamical phason variables in the lowest order.
The Fourier transform with respect to relative coordinates is given by
\begin{align}
\chi^{\mu\nu}(\bm q,\omega) &\simeq
 \frac{S^2}{Q^2}\sum_{ij}Q^\mu_i Q^\nu_j
\,
\overline{
\la \eta_i (\bm q,\omega) \eta_j (-\bm q,-\omega) \ra
}
\nonumber \\
&\hspace{-15mm} + \frac{S^2}{4}\sum_{i\sg} n_{i\sg}^\mu n_{i,-\sg}^\nu 
\,
\overline{
\la \phi_i(\bm q-\sg\bm Q_i,\omega) \phi_i(-\bm q+\sg\bm Q_i,-\omega) \ra
}
\end{align}
where $\mu,\nu=x,y,z$ and $\bm n_{i\sg}=\hat {\bm z} - \imu \sg \hat {\bm Q}_i \times \hat {\bm z}$ with $\sg=\pm 1$.
The spatially fast oscillating part has been dropped and $\omega>0$ is assumed.
We also define the quasi-local magnetic response function by 
$\chi^{\mu\nu}_{\rm loc}(\omega) = \Omega^{-1}\sum_{\bm q}\chi^{\mu\nu}(\bm q,\omega)$ with $\Omega$ being a volume of the system, which has the form
\begin{align}
\chi^{\mu\nu}_{\rm loc} (\omega) \simeq
\delta_{\mu\nu}(\delta_{\mu x}+\delta_{\mu y}+2\delta_{\mu z})
\frac {\hbar a^d S^2 }{2\Omega}
\sum_{\bm q} {\mathscr{G}}_{xx}(\bm q, \omega)
.
\end{align}
Since the relation $\eta \sim \omega \phi$ holds from the canonical conjugate relation for a dynamical part, we have neglected the $\eta$-field contributions which are small at low frequencies.
Neglecting also the $q_z$-linear term that appears as a higher-order contribution, we obtain the imaginary part of local magnetic susceptibility for three dimensions as
\begin{align}
&\imag \chi_{\rm loc}^{zz}(\omega) \simeq
\frac{\hbar^2 \Sigma_2 \omega}{8\pi J^{3/2} S \sqrt{\Sigma_1}}
,
\end{align}
 and for two dimensions 
\begin{align}
\imag \chi_{\rm loc}^{{\rm (2D)} zz}(\omega)
& \simeq \frac{\hbar^2 \Sigma_2 \omega}{4\pi J\Sigma_1}
.
\end{align}
With these expressions the local magnetic relaxation rate is derived as
\begin{align}
\frac 1 {T_1T} =
\frac{A^2k_{\rm B}}{\hbar^3}
\lim_{\omega\rightarrow 0} \frac{\imag \chi_{\rm loc}^{zz}(\omega)}{\omega}
, \label{eq:relation_rate}
\end{align}
where $A$ is a `hyperfine' interaction parameter which has the dimension of energy.
The detailed expressions for local magnetic relaxation rate are summarized in Tab.~\ref{tab:expr}.
When the disorder effect becomes weaker, the magnetic relaxation rate is increased because of the more low-energy state density of phasons that are involved in the relaxation process.
The typical frequency dependences of local magnetic spectrum for three and two dimensional SkX are also shown in Figs.~\ref{fig:plots}(a) and (b), which have peaked structures around $\omega=\omega_{\rm p}$ reflecting a resonance at pinning frequency (see Sec.~\ref{sec:discussion} for a choice of parameters).

\subsection{Dynamical coupling to uniform external fields}

We consider the coupling between skyrmion spin texture and external electromagnetic fields.
This can be obtained in a manner similar to the discussion for CDW, where the dynamics of phasons carry the electrical current \cite{lee74}.
For a triple helix structure of SkX, we superpose the results for the single helix case which are derived by picking up the lowest-order electron bubble contribution (see Appendix \ref{sec:appendix2}).
We then obtain the following couplings with uniform electric and magnetic fields in the action:
\begin{align}
\mathscr S_{\rm ext} &= \int \diff t\diff \bm r \left[
\frac{e\hbar^2 \Delta^2 N''(0) Q}{\sqrt 6 m a^d}  \bm \phi \cdot \bm E
- \frac{\sqrt 6 g_s e\hbar S}{4ma^d} \bm \eta \cdot \bm B
\right]
, \label{eq:coupling_EB}
\end{align}
where we have defined the vectors $\bm \phi=(\phi_x,\phi_y,0)$ and $\bm \eta =(\eta_x,\eta_y,0)$.
The mean-field for electrons is given by $\Delta = \frac 1 2 J_{\rm H}S$ with $J_{\rm H}$
being a Hund coupling from Coulomb repulsive interaction, and $N(0)$ is the density of states at Fermi level.
We note the relation $N''(0)\sim \ep_{\rm F}^{-3}$ where $\ep_{\rm F}\sim \frac{\hbar^2}{ma^2}$ is the Fermi energy.

The electrical conductivity $\sg_{\al\beta}$ is defined by the relation $j_{\al}=\sum_{\beta}\sg_{\al\beta} E_{\beta}$, and is given by
\begin{align}
\sigma_{\al\beta} (\omega) &= a^d \left( \frac{e\hbar^2 \Delta^2 N''(0)Q}{\sqrt 6 ma^d} \right)^2 \imu \omega 
{\mathscr G}_{\al\beta}(\bm 0, \omega)
. \label{eq:conductivity}
\end{align}
The typical frequency dependences for $d=3$ and $d=2$ are plotted in Figs.~\ref{fig:plots}(e) and (f), respectively (see Sec.~\ref{sec:discussion} for a choice of parameters).
The spectrum has a peaked structure at $\omega \simeq \omega_{\rm p}$, and the conductivity becomes zero when the frequency approaches to zero.
This is a typical behavior of the pinned state of elastic media \cite{fukuyama78, chitra01}.
It is characteristic for $d=2$ that the two resonance peak structures can be seen in the dynamical spectral functions as shown in Fig.~\ref{fig:plots}(f). 
This is because the pinning frequency $\omega_{\rm p}$ is comparable to the higher energy branch ($\sim D^2S/J$) in dispersion relation shown in Fig.~\ref{fig:dispersion}.
In contrast, the two energy scales are much separated in the three dimensional SkX as shown in Fig.~\ref{fig:plots}(e), since the pinning frequency in $d=3$ is much smaller than in $d=2$ as discussed before.
We note that the low-energy properties for SkX (and also for HMS) are gapless since the conductivity behaves as $\sigma(\omega)\propto \omega^2$.
This is similar to the behavior of the Bose glass \cite{nelson93}.

Next we consider the magnetic response functions.
The dynamical susceptibility defined by $M_{\al} = \sum_{\beta} \chi_{\al\beta} B_{\beta}$, where $\bm M$ is the magnetization per unit volume, is given by
\begin{align}
\chi_{\al\beta}(\omega) &= a^d
\left(
\frac{\sqrt 6 g_s e\hbar S}{4ma^d}
\right)^2
{\mathscr F}_{\al\beta} (\bm 0, \omega)
, \label{eq:magnetic_sus}
\end{align}
where $\mathscr F$ is defined in Eq.~\eqref{eq:Green_beta}.
We can also have the cross term representing the magnetoelectric effect: $M_{\al} = \chi^{\rm ME}_{\al\beta} E_{\beta}$.
The response kernel has the form
\begin{align}
\chi_{\al\beta}^{\rm ME}(\omega) &= 
\frac{3g_s e^2\hbar^4 \Delta^2N''(0)}{8 m^2D a^{2d-2}}
\imu\omega {\mathscr G}_{\al\beta}(\bm 0, \omega)
. \label{eq:magnetoelectric}
\end{align}
These response functions ($\sg_{\al\beta}$, $\chi_{\al\beta}$, $\chi^{\rm ME}_{\al\beta}$) at $\omega=\omega_{\rm p}$ in the weak disorder limit are listed in Tab.~\ref{tab:expr}.
While the pinning frequency is sensitively dependent on the dimension of system, the response functions at resonance frequencies are not the case.
This insensitivity is characteristic for SkX, 
where the $\omega$-linear term appears in Green functions due to their topological properties.
In contrast,
the values are dependent on the dimensionality for HMS (see Tab.~\ref{tab:expr2}).
A further comparison between SkX and HMS is given in Sec.~\ref{sec:discussion}.

\subsection{Directional dichroism in three dimensional SkX}

We can also show that a directional dichroism, where the responses to external electromagnetic waves propagating in the directions $+\bm q$ and $-\bm q$ are different, appears along the magnetic field direction (parallel to $\hat{\bm z}$) for the three dimensional SkX.
In the realistic situation, since the conductivity from conduction electrons is much larger than the one from phason contribution, and also the electromagnetic field is reflected and cannot penetrate through the sample, we expect the possible observation in insulating compounds.

Noting the relations $\bm E = -\partial_t \bm A$ and $\bm B = \bm \nabla \times \bm A$, the dynamical coupling term \eqref{eq:coupling_EB} can be rewritten in terms of vector potential.
We then obtain the current in the form $j_{\mu} = \sum_{\nu} \Pi_{\mu\nu} A_\nu$ with $\mu,\nu=x,y,z$.
For $\bm q= (0,0,q_z)$, i.e. the light 
propagating along $z$-direction, only the kernels $\Pi_{xx},\Pi_{yy},\Pi_{xy},\Pi_{yx}$ are finite and are given by the following $2\times 2$ matrix:
\begin{widetext}
\begin{align}
\hat \Pi(q_z,\omega) = a^d \left( \frac{\sqrt 6 g_se\hbar S}{4 ma^d} \right)^2 q_z^2 \hat g(q_z)
+  a^d \left[
\frac{e\hbar^2 \Delta^2 N''(0) Q}{\sqrt 6 m a^d} \hat 1 + \frac{3\sqrt 6 g_se\hbar^2 S^2}{16ma^d} \imu q_z \hat g(q_z) \hat \epsilon
\right]^2 \omega^2 \hat {\mathscr G} (q_z,\omega)
,
\end{align}
\end{widetext}
which produces the difference between the responses for $+q_z$ and $-q_z$.
Let us estimate the magnitudes of the directional dichroism at the resonance frequencies where the response function becomes maximum.
We use the relation $\omega = c |q_z|$ with $c$ being the speed of light and assume that the polarization of light is along $x$-direction.
We can then write down the ratio between the energy absorption coefficients for the lights having the wave vectors $+q_z$ and $-q_z$ as
\begin{align}
&\frac{\imag \Pi_{xx}(-\omega/c,\omega)}{\imag \Pi_{xx}(+\omega/c,\omega)}
\equiv 
\frac{1+r(\omega)}{1-r(\omega)}
.\label{eq:intens_ratio}
\end{align}
The effect of directional dichroism is maximized when $r\sim 1$.
For a typical value of $\omega$, we choose $\omega=\omega_{\rm p}$ and $\omega=\frac{D^2S}{J\hbar}$ which correspond respectively to the characteristic energies for lower ($\omega_{\rm low}$) and higher ($\omega_{\rm high}$) branches in the dispersion relations shown in Fig.~\ref{fig:dispersion}.
The quantity $r$ for each case is estimated as
\begin{align}
r_{\rm low}&\equiv r(\omega_{\rm p}) \sim \frac{V_{\rm imp}^4}{S D^3J}\left( \frac{a}{\ell} \right)^6 
 \left( \frac{\ep_{\rm F}}{J_{\rm H}} \right)^2 \frac{v_{\rm F}}{c}
, \label{eq:r1}\\
r_{\rm high}&\equiv r(\tfrac{D^2S}{J\hbar}) \sim \frac{J}{S D}
 \left( \frac{\ep_{\rm F}}{J_{\rm H}} \right)^2 \frac{v_{\rm F}}{c}
. \label{eq:r2}
\end{align}
where we have used the expressions for the Fermi energy $ \ep_{\rm F}\sim \tfrac{\hbar^2 }{ma^2}$ and Fermi velocity $v_{\rm F}\sim \tfrac{\hbar}{ma}$ for an order estimation.
The value of $r_{\rm low}$ should be much smaller than the unity because of the factor $(a/\ell)^6$.
Hence the directional dichroism is more likely to be observed in the higher-energy $\omega_{\rm high}$ branch.
Let us also estimate the value of the absorption coefficient. 
The intensity $I$ of light decays as $I=I_0 \epn^{-\al x}$ with $x$ being a distance that the light travels inside the material. 
The coefficient $\al$ for both $\omega_{\rm low}$ and $\omega_{\rm high}$ branches is given by
\begin{align}
\al = \frac{\real \sg_{xx}(\tfrac{D^2S}{J\hbar})}{c\epsilon_0}
\sim \frac{S^3}{a } \left( \frac{e^2}{4\pi \epsilon_0\hbar c} \right) \left( \frac{J_{\rm H}}{ \ep_{\rm F}} \right)^4 \left( \frac{D}{J} \right)^2
, \label{eq:absorption}
\end{align}
where $\epsilon_0$ is a permittivity of vacuum.
As shown in the next section, a moderate absorption coefficient $\al$ can be obtained for three dimensional SkX.

If we take $\bm q\perp \hat {\bm z}$, the directional dichroism does not appear in our model.
On the other hand, it has been suggested that such an effect can be detected by the light propagating along the $xy$-plane \cite{mochizuki13} by assuming the cubic symmetry of the crystal.
However, this material-specific effect is not included in the present model.
Our directional dichroism discussed here is rather related to the general feature of the three dimensional SkX, and is not specific to a particular material.

\begin{table*}
\begin{tabular}{c||c|c|c|c|c|c|c}
\hline
{\it SkX} & 
$\omega_{\rm p}$ & 
$\xi_{\rm p}$ & 
$1/T_1 T$ & 
$\real \sg (\omega_{\rm p})$ & 
$\real \chi (\omega_{\rm p})$ & 
$\real \chi^{\rm ME} (\omega_{\rm p})$ & 
$j_{\rm c}$ 
\\
\hline
\parbox[c][1cm][c]{0cm}{}
${d=3}$ & 
$\dfrac{V_{\rm imp}^4 S^3 a^6}{\hbar J^3 \ell^6}$ &
$\dfrac{J^2 \ell^3}{V_{\rm imp}^2 a^2}$ &
$\dfrac{A^2 k_{\rm B} \ell^3}{\hbar V_{\rm imp}^2 S a^3}$ &
$\dfrac{e^2S^3 J_{\rm H}^4D^2}{\hbar a \ep_{\rm F}^4J^2}$ &
$\dfrac{e^2a\ep_{\rm F}^2J}{\hbar^2D^2}$ &
$\dfrac{e^2 S J_{\rm H}^2}{\hbar \ep_{\rm F}D}$ &
$\dfrac{|e|Sa^4 V_{\rm imp}^4}{\hbar \ell^6 D J^2 }$ 
\\
\hline
\parbox[c][1cm][c]{0cm}{}
${d=2}$ &
$\dfrac{V_{\rm imp}^2S^3 a^2}{\hbar J\ell^2}$&
$\dfrac{J\ell}{V_{\rm imp}}$&
$\dfrac{A^2 k_{\rm B} \ell^2}{\hbar V_{\rm imp}^2Sa^2}$ &
$\dfrac{e^2S^3 J_{\rm H}^4D^2}{\hbar \ep_{\rm F}^4J^2}$ &
$\dfrac{e^2a^2\ep_{\rm F}^2J}{\hbar^2D^2}$ &
$\dfrac{e^2 S aJ_{\rm H}^2}{\hbar \ep_{\rm F}D}$ &
$\dfrac{|e|S V_{\rm imp}^2}{\hbar \ell^2 D }$
\\
\hline 
\parbox[c][1cm][c]{0cm}{}
ratio &
$\left(\dfrac{V_{\rm imp}}{J}\right)^2 \left(\dfrac{\ell}{a}\right)^{-4}$ &
$\left(\dfrac{V_{\rm imp}}{J}\right)^{-1} \left(\dfrac{\ell}{a}\right)^2$ &
$\dfrac{\ell}{a}$&
$a^{-1}$&
$a^{-1}$&
$a^{-1}$&
$\left(\dfrac{V_{\rm imp}}{J}\right)^{2}\left(\dfrac{\ell}{a}\right)^{-4}$
\\
\hline 
\end{tabular}
\caption{
Dependencies of physical quantities on the parameters
($J, D, a, S, V_{\rm imp}, \ell,  \ep_{\rm F}, J_{\rm H}$) included in the theory of the disordered SkX.
Here we omit the constant factors.
The results for the clean limit without disorders can be derived by taking the limit $V_{\rm imp}\rightarrow 0$ or $\ell \rightarrow \infty$.
}
\label{tab:expr}
\end{table*}

\begin{table*}
\begin{center}
\begin{tabular}{c||c|c|c|c|c}
\hline
 {\it HMS} & 
$\omega_{\rm p}$ & 
$1/T_1 T$&
$\real \sg(\omega_{\rm p})$ &
$\real \chi (\omega_{\rm p})$ & 
$\real \chi^{\rm ME} (\omega_{\rm p})$ 
\\
\hline
\parbox[c][1cm][c]{0cm}{}
${d=3}$ & 
$\dfrac{DV_{\rm imp}^2Sa^3}{\hbar J^2\ell^3}$ &
$\dfrac{A^2k_{\rm B}}{\hbar DJS}$&
$\dfrac{e^2S^3\ell^3J_{\rm H}^4D^3}{\hbar a^4\ep_{\rm F}^4J V_{\rm imp}^2}$&
$\dfrac{e^2a\ep_{\rm F}^2J}{\hbar^2D^2}$&
$\dfrac{e^2 S \ell^3J_{\rm H}^2J}{\hbar a^3 V_{\rm imp}^2\ep_{\rm F}}$
\\
\hline
\parbox[c][1cm][c]{0cm}{}
${d=2}$ &
$\dfrac{DV_{\rm imp}Sa}{\hbar J\ell}$&
$\dfrac{A^2k_{\rm B}\ell}{\hbar V_{\rm imp}SDa}$&
$\dfrac{e^2S^3\ell J_{\rm H}^4D^3}{\hbar a \ep_{\rm F}^4V_{\rm imp}J^2}$&
$\dfrac{e^2a^2\ep_{\rm F}^2J}{\hbar^2D^2}$&
$\dfrac{e^2 S \ell J_{\rm H}^2}{\hbar \ep_{\rm F} V_{\rm imp}}$
\\
\hline 
\parbox[c][1cm][c]{0cm}{}
ratio &
$\dfrac{V_{\rm imp}}{J} \left( \dfrac{\ell}{a} \right)^{-2}$ &
$\dfrac{V_{\rm imp}}{J} \left( \dfrac{\ell}{a} \right)^{-1}$ &
$\left(\dfrac{V_{\rm imp}}{J}\right)^{-1}\left( \dfrac{\ell}{a} \right)^{2}a^{-1}$ &
$a^{-1}$ &
$\left(\dfrac{V_{\rm imp}}{J}\right)^{-1} \left( \dfrac{\ell}{a} \right)^{2}a^{-1}$ 
\\
\hline 
\end{tabular}
\end{center}
\caption{
Analogous table to Tab.~\ref{tab:expr} but here for helimagnets.
The expression for $\xi_{\rm p}$ in helimagnets is same as that in SkX and is not shown.
}
\label{tab:expr2}
\end{table*}

\begin{table*}
\begin{center}
\begin{tabular}{c||c|c|c|c|c|c}
\hline
{\it SkX} / {\it HMS} & 
$\omega_{\rm p}$ & 
$\xi_{\rm p}$ & 
$1/T_1 T$ & 
$\real \sg (\omega_{\rm p})$ & 
$\real \chi (\omega_{\rm p})$ & 
$\real \chi^{\rm ME} (\omega_{\rm p})$ 
\\
\hline
\parbox[c][1cm][c]{0cm}{}
${d=3}$ & 
$\dfrac{V_{\rm imp}^2S^2 a^3}{J^2\ell^3}$ &
$1$ &
$\dfrac{DJ\ell^3}{V_{\rm imp}^2a^3}$ &
$\dfrac{V_{\rm imp}^2  a^3}{DJ \ell^3}$ &
$1$ &
$\dfrac{V_{\rm imp}^2a^3}{DJ\ell^3}$
\\
\hline
\parbox[c][1cm][c]{0cm}{}
${d=2}$ &
$\dfrac{V_{\rm imp}S^2 a}{J\ell}$ &
$1$ &
$\dfrac{D \ell}{V_{\rm imp} a}$ &
$\dfrac{V_{\rm imp}a}{D\ell}$ &
$1$ &
$\dfrac{V_{\rm imp}a}{D\ell}$
\\
\hline 
\end{tabular}
\end{center}
\caption{
Ratio between physical quantities for SkX and HMS listed in Tabs. \ref{tab:expr} and \ref{tab:expr2}.
}
\label{tab:expr3}
\end{table*}


\section{Discussion} \label{sec:discussion}

As discussed above, we have clarified the characteristic properties for SkX in the presence of disorders.
In order to make a closer connection with real materials, it is useful to estimate typical values of the parameters and physical quantities.
Let us first 
consider the distance $\ell$ between impurities, which can be estimated from the critical current density $j_{\rm c}$ for SkX realized in metallic compounds such as MnSi.
To do this, we consider the energy density for spin-transfer torque introduced by external current \cite{zang11,tatara14}:
\begin{align}
\mathscr{H}_{\rm ext} 
&= \frac{\mathcal B}{Q} \int \frac{\diff \bm r}{a^{d-3}}
\sum_{\al\beta} \epsilon_{\al\beta} (Sj_{\rm ext,\al}) \phi_{\beta}
\end{align}
We have introduced the uniform emergent magnetic field $\mathcal B \sim Q^2 \hbar / |e|$, which corresponds to the number density of skyrmions.
Within this expression, the anisotropy for $j_{\rm c}$ does not appear.
At the critical current density $j=j_{\rm c}$, this energy is put equal to the pinning energy given in the second term of the right-hand side in Eq.~\eqref{eq:energy_pin}.
This is similar to the estimation for the electric-field depinning of CDW with weak impurities \cite{lee79}.
We obtain the relation
\begin{align}
\frac{\mathcal B S j_{\rm c}}{Q a^{d-3}} \sim 
\frac{V_{\rm imp}S^2}{(\xi_{\rm p} \ell)^{d/2} }
. \label{eq:critical_cur}
\end{align}
Based on this, the expressions of critical current densities for three and two dimensions are obtained and are listed in Tab.~\ref{tab:expr}.
As in the pinning frequency, the critical current density for $d=3$ also becomes much smaller than the one in $d=2$.
With increasing the size of skyrmions, i.e. $Q\rightarrow 0$, the energy supply from the external current decreases because of $\mathcal B \propto Q^2$ as seen in Eq.~\eqref{eq:critical_cur}, which leads to the larger critical current density.
This is in sharp contrast with the ordinary density waves, where the critical current density becomes smaller when the modulation period increases since the driving force is not dependent on the period.

The relation in Eq.~\eqref{eq:critical_cur} is used to estimate the distance $\ell$ between impurities.
For the SkX phase in the three dimensional material MnSi, we take $J\simeq 13 {\rm \,meV}$, $D/J = 0.1$ \cite{grigoriev06,ishikawa77} 
and $S=0.1$.
The characteristic energy scale $\omega_{\rm high}$ 
for the higher-energy phason mode
 is given by $D^2S/J$ (see Fig.~\ref{fig:dispersion}), 
whose relevant frequency is $\sim 1 {\rm \,GHz}$.
We also use 
the lattice constant 
$a=2.9 {\rm \, \AA}$ \cite{ishikawa77}, and assume $V_{\rm imp}\simeq J$ for the impurity potential.
With the experimental critical current density
$j_{\rm c} = 10^6 {\rm \,A/m^2}$ \cite{jonietz10,schulz12},
we obtain $\ell \sim 1 {\rm \,nm}$.
The pinning frequency and length in three dimensional SkX are then 
$\omega_{\rm p}/2\pi \sim 10^{6} {\rm \,Hz}$ 
and 
$\xi_{\rm p} \sim 10^{3} {\rm \,nm}$, respectively,
indicating that the system is well located in 
the collective or weak pinning regime ($\xi_{\rm p}/\ell \sim 10^3$).
For two dimensions, the dependences of pinning frequency on the potential $V_{\rm imp}$ and the distance $\ell$ is weaker compared to the $d=3$ case, which results in the shorter pinning length $\xi_{\rm p}/\ell \sim 10^{1}$.
Hence the two-dimensional SkX is nearly located at the edge of the region where the weak-pinning approximation can be applied.
The shorter pinning length corresponds to the larger pinning frequency which is the order of $\omega_{\rm p}/2\pi\sim 10^9 \,{\rm Hz}$ comparable to $D^2S/\hbar J$.
This is the reason why we see the two resonance peaks in $\sigma(\omega)$ with a single window as shown in Fig.~\ref{fig:plots}(f).

As emphasized in the introduction, it is characteristic for chiral magnets that the physical quantities in SkX and HMS can be directly compared, which is not the case for the other disordered systems such as CDW, vortex lattice and Wigner crystals.
This 
situation leads us to a deeper understanding of the pinned elastic media.
We summarize the results for HMS in Tab.~\ref{tab:expr2} (see Appendix \ref{sec:appendix1} for derivation).
The dynamical quantities (local magnetic spectrum and conductivities) are also plotted in Figs.~\ref{fig:plots}(c), (d), (g), (h), where the single peak appears in the spectrum as is different from SkX.
This is because there is only one phason field variable and the dispersion relation has only one branch as shown in the dotted line in the left panel of Fig.~\ref{fig:dispersion}.
It is characteristic for three dimensional HMS that there is no $\omega$ dependence in local magnetic spectrum as shown in Fig.~\ref{fig:plots}(c), which is related to the fact that the HMS in $d=3$ is located at the critically damped case with $\zeta=1$ in our model.

The ratios between the quantities in SkX and HMS are also given in Tab.~\ref{tab:expr3}.
First let us look at the pinning frequency in Tab.~\ref{tab:expr3} for $d=3$.
Since the ratio is given by $\omega_{\rm p}^{\rm SkX}/\omega_{\rm p}^{\rm HMS}\propto (V_{\rm imp}/J)^2 (a/\ell)^3$ with the relations $a \ll \ell$ and $V_{\rm imp}\sim J$, the pinning frequency is substantially reduced once we enter the SkX phase from HMS.
This change is a consequence of the gyro dynamics that appears only in SkX.
In contrast, the pinning length is determined by the statics in the limit $\omega\rightarrow 0$, and is nearly unaffected except for a constant factor.
In $d=2$ the ratio is $\omega_{\rm p}^{\rm SkX}/\omega_{\rm p}^{\rm HMS}\propto (V_{\rm imp}/J)^1 (a/\ell)^1$, which also suggests the reduced pinning frequency in SkX, but the difference is much milder than the $d=3$ case.

The change in dynamics between SkX and HMS is reflected also in different local magnetic responses.
For SkX, the local magnetic relaxation rate has the form  $1/T_1 \propto V_{\rm imp}^{-2}\ell^3$ for $d=3$ and $\propto V_{\rm imp}^{-2}\ell^2$ for $d=2$. 
On the other hand, $1/T_1$ for HMS has a weaker dependence on the impurity profiles, and notably the HMS in $d=3$ has no dependence on impurities. 
This results in an enhanced local magnetic relaxation rate $1/T_1$ in SkX compared to HMS.
The enhancement of $1/T_1$ originates from the difference of the low-lying energy modes which are $\omega\sim \bm q^2$ for SkX and $\omega\sim|\bm q|$ for HMS, where the quadratic dispersion creates the more low-energy states.

According to Tab.~\ref{tab:expr3}, for ac conductivity ($\sg$) and magnetoelectric susceptibility ($\chi^{\rm ME}$), the ratio between SkX and HMS are sensitively dependent on the impurity profiles, while the magnetic susceptibility ($\chi$) is not. This is because the phason field $\phi$ is directly involved in the expression of $\sg$ and $\chi^{\rm ME}$ as in Eqs.~\eqref{eq:conductivity} and \eqref{eq:magnetoelectric}, and the low-energy dynamics is much modified by impurities.
For $\chi$, on the other hand, the $\eta$-field conjugate to $\phi$ is instead relevant, whose Green function is given by Eq.~\eqref{eq:Green_beta} and has the constant part $\hat g$. 
Because of the presence of this static susceptibility, the low-energy dynamics is not sensitive to the weak disorders and there is not much difference between SkX and HMS.

We have also proposed 
the 
possible observation of directional dichroism for three dimensional SkX. By assuming the magnitude relations $c/v_{\rm F}\sim 10^2$, $D/J\sim 10^{-1}$ and $J_{\rm H}/\ep_{\rm F} \sim 1 $ in Eqs.~\eqref{eq:intens_ratio}, \eqref{eq:r1} and \eqref{eq:r2}, we get $r_{\rm high}\sim 1$ and expect that the directional dichroism is possibly observed in the higher-energy $\omega_{\rm high}$ branch. As for the absorption coefficient $\al$ defined by Eq.~\eqref{eq:absorption}, it is estimated as $\al \sim 10^{-7} a^{-1}$. Hence the sample thickness $t\sim 1 {\rm \ mm}$ that gives $\al t\sim 1$ is appropriate for detecting the directional dichroism along the magnetic field direction.

\section{Summary}

We have investigated the dynamical properties of the disordered magnetic skyrmions at low energies by using the replica field theory combined with the Gaussian variational approximation.
The Green functions for the phasons, which describes the low-frequency and long-wavelength behaviors, have been explicitly calculated.
With these quantities, we have obtained the pinning frequency and length specific to the pinned skyrmion states.
The magnetic and electric response functions have been derived in two and three dimensional skyrmion crystals, and 
compared with the results of the topologically trivial helimagnetic state to illuminate the characteristic properties of magnetic skyrmions in the presence of disorders.
The difference of the low-energy glassy properties between these two states is prominent, and hence the chiral magnets are regarded as a unique system to show the topological phase transition among the glassy states by tuning temperature or applied magnetic field, which provide a deeper understanding for physics of disorders.
These 
features of disordered magnetic skyrmions should give a clear distinction from the other physical systems previously discussed.

Specifically for
the three dimensional skyrmion crystal, we have 
clarified the nonreciprocal nature of collective excitation modes and their electromagnetic responses along the direction of external magnetic field.
The nonreciprocal properties of chiral magnets put forward a new point of view for the elastic media, and can be further investigated in the other related systems such as three dimensional vortex lattices in noncentrosymmetric superconductors.
Our results for pinned regime of skyrmions can also be a foundation for further exploring the glassy states and nonreciprocal responses in the moving regime under the electronic current.

\vspace{10mm}

\section*{Acknowledgement}
The authors thank 
F. Kagawa, W. Koshibae, S. Seki, T. Yokouchi,
M. Kawasaki and Y. Tokura for fruitful discussions, and M. Ishida for making Fig.~\ref{fig:illust}. 
This work was supported by 
JSPS Grant-in-Aid for Scientific Research
No. 16H04021 (SH), No. 24224009 and No. 26103006 (NN) from MEXT, Japan, and ImPACT Program
of Council for Science, Technology and Innovation (Cabinet
office, Government of Japan) (NN). 
This work was also supported by CREST, JST.

\appendix
\section{Derivation of effective action for skyrmion crystal} \label{sec:appendixNEW}

Here let us derive the effective action for the disordered SkX.
The effect from pinning potentials
is incorporated by using a replica trick \cite{mezard91,giamarchi94,giamarchi95,giamarchi96}.
Within this theoretical framework, the free energy after averaging over impurity configuration is evaluated by using the relation
$\overline{\ln Z} 
= \lim_{n\to 0} \frac 1 n \ln \overline{Z^n}$ for $n\in \mathbb Z$.
The effective action for the $n$-replicated system is written as
\begin{widetext}
\begin{align}
&\hspace{-3mm}\mathscr{S}_{\rm eff}
= \sum_{\mathrm a=1}^{n}\frac{1}{a^d} \sum_{\bm q\al\beta} \int \diff \tau
\left[\frac{9\imu \hbar S }{8} \epsilon_{\al\beta}
 \phi_{\mathrm a\al} (\bm q,\tau)
\dot \phi_{\mathrm a\beta}(-\bm q,\tau)
+
\frac{JS^2a^2\bm q^2}{2}\delta_{\al\beta}
\phi_{\mathrm a\al}(\bm q,\tau)
\phi_{\mathrm a\al} (-\bm q,\tau)
\right.
 \nonumber \\
&\hspace{35mm}
+ \dot \phi_{\mathrm a\al} (\bm q,\tau)
\left.
\left(
\frac{9\hbar^2 J}{8D^2} \delta_{\al\beta}  -  \frac{27 \imu \hbar^2 J^2 a q_z}{8D^3} \epsilon_{\al\beta}
\right)
\dot \phi_{\mathrm a\beta}(-\bm q,\tau)
\right] 
\nonumber \\
&\hspace{8mm} - \frac{V_{\rm imp}^2S^4 a^d}{4\hbar \ell^d}\sum_{\mathrm a \mathrm b}
\int \frac{\diff \bm r}{a^d}\int \diff \tau\diff\tau'
\left[\sum_i\cos\left(\phi_{\mathrm ai}(\bm r,\tau) - \phi_{\mathrm bi}(\bm r,\tau')\right) \right]^2
, \label{eq:action_replica}
\end{align}
\end{widetext}
which describes the low-energy physics of disordered phasons.
For analysis, we take the Gaussian variational approximation \cite{mezard91,giamarchi96}, in which the system is described by the action
\begin{align}
\mathscr{S}_{0} &=
\frac{1}{2a^d}
\sum_{\mathrm a \mathrm b}
\sum_{\bm qn}\sum_{\al\beta} 
\nonumber \\
&\ \ \times
\phi_{\mathrm a\al}(\bm q,\omega_n) \, 
[G^{-1}(\bm q,\imu \omega_n)]_{\beta\al}^{\mathrm b \mathrm a} \,
\phi_{\mathrm b\beta}(-\bm q,-\imu \omega_n)
\end{align}
and the variational 
functional is also defined as
$\mathscr{S}_{\rm var} = -\hbar \ln Z_{0} 
+ \la \mathscr{S}_{\rm eff} - \mathscr{S}_{0} \ra_{0} $ 
where $Z_{0} = \int \mathscr{D}\{\phi_{\al}\} \exp[-\mathscr{S}_{0}/\hbar]$.
The bosonic Matsubara frequency is given by $\hbar \omega_n = 2\pi n k_{\rm B}T$.
We introduce the self-energy from impurities by the relation
$
(G^{-1})_{\al\beta}^{\mathrm a \mathrm b} = (G_{V=0}^{-1})_{\al\beta}^{\mathrm a \mathrm b} - \sigma_{\al\beta}^{\mathrm a \mathrm b}
$, and the stationary condition of variational functional with respect to the self-energy $\sg$ gives a set of equations to determine the Green function.
Finally, the Green function of the original system is given by
$\mathscr G_{\al\beta} = \lim_{n\rightarrow 0} \frac{1}{n} \sum_{\mathrm a} G_{\al\beta}^{\mathrm a \mathrm a}$.

Physically $n$ replicas represent many metastable states induced by impurities, and have similar properties: $G^{\mathrm a \mathrm a} = G_0$. 
As for the off-diagonal elements of the Green function, one possible situation is that these replicas are symmetric and the correlation functions between replicas are not dependent on replica index: $G^{\mathrm a \mathrm b} = G_1$ for $\mathrm a \neq \mathrm b$ (replica symmetric solution). However, it has been known that this solution is unstable for two and three dimensions \cite{mezard91}.
Hence we need to consider the replica symmetry broken solution. 
By assuming the hierarchical replica symmetry breaking ansatz \cite{mezard_book}, the off-diagonal components can be labeled by the one parameter $u\in[0,1]$. 
This procedure is proven to be exact in the Sherrington-Kirkpatrick model for a spin glass \cite{talagrand06}.

The self-energy can be written in the form
$\sg (\imu \omega_n) = \Sigma_1(1-\delta_{n0}) + I(\imu \omega_n)$
with $I(\imu \omega_n\rightarrow 0)= \Sigma_2 \hbar |\omega_n|$, 
which
is finite only for the diagonal component \cite{chitra01}.
Assuming the full replica symmetry breaking solution, we obtain an explicit form of the self-energy after straightforward calculation similar to Refs.~\cite{giamarchi96} and \cite{chitra01}.
The equations that determine $\Sigma_1$ and $I(\imu\omega_n)$ are derived as
\begin{align}
1 &= \frac{6V_{\rm imp}^2S^4 a^{2d}}{\ell^d \Omega} \sum_{\bm q} \frac{1}{[G^0_{\al\al}(\bm q,0) + \Sigma_1]^2}
, \\
I(\imu\omega_n ) &= \frac{6V_{\rm imp}^2S^4 a^{2d}}{\ell^d \Omega} 
\sum_{\bm q} \left[ \frac{1}{G^0_{\al\al}(\bm q,0) + \Sigma_1} 
- \mathscr G_{\al\al}(\bm q,\imu\omega_n) \right]
,
\end{align}
where $\hat G^0 (\bm q,\imu\omega_n)$ is the free phason Green function without disorder effects, and $\mathscr G_{\al\beta}=\la \phi_{\al} \phi_{\beta} \ra/a^d\hbar$ is the full version given by
$\hat {\mathscr G}^{-1}(\bm q,\imu \omega_n) = [\hat G^0(\bm q,\imu \omega_n)]^{-1} - \sg(\imu\omega_n)\hat 1$.
In the quantum theory of (2+1) dimensional skyrmion system, i.e. the $d=2$ case in our model, the response function has a static self-energy $\Sigma_1$, which reflects the localization of skyrmions as in the case of the Bose glass \cite{giamarchi96, nelson93}.

By performing the analytic continuation $\imu \omega_n \rightarrow \omega$, the real-frequency Green function is written in the $2\times 2$ matrix form, which is explicitly written in Eq.~\eqref{eq:green} of the main text.
Since the Green functions are derived with the Matsubara formalism, the results can be applied at finite temperatures.
While the functional form of real-frequency Green functions after the analytic continuation does not depend on temperature,
we note that
the temperature dependence enters through the parameters such as a modulus $S$ of the spin moment.

\vspace{7mm}

\section{Results for helimagnetic state} \label{sec:appendix1}

We here summarize the properties of HMS, which are compared to SkX.
The spin configuration for the simple HMS is given by
\begin{align}
\bm S &= S\eta \hat{\bm y} + S\sqrt{1-\eta^2}\left[
 \hat {\bm z} \cos (Qy+\phi)
+\hat {\bm x} \sin (Qy+\phi)
\right]
,
\end{align}
where the modulation vector is along $y$-direction ($\hat {\bm Q}=\hat {\bm y}$).
After integrating out the $\eta$-field, we obtain the effective low-energy action for phason replica fields as
\begin{align}
&\mathscr S_{\rm eff} = \sum_{\mathrm a} \int \diff \tau \int \frac{\diff \bm r}{a^d} \left[
\frac{JS^2a^2}{2} (\bm \nabla \phi_{\mathrm a})^2
+\frac{J\hbar^2}{2D^2} \dot \phi_{\mathrm a} ^2
\right]
\nonumber \\
&\hspace{1mm}  - \frac{V_{\rm imp}^2S^4 a^d}{4\hbar \ell^d}\sum_{\mathrm a \mathrm b}
\int \frac{\diff \bm r}{a^d}\int \diff \tau \diff\tau'
\cos^2 \left( \phi_{\mathrm a}(\bm r,\tau) - \phi_{\mathrm b}(\bm r,\tau') \right)
.
\end{align}
Tracing the same procedure given in the SkX case, we obtain the Green function 
\begin{align}
\mathscr G(\bm q, \omega)^{-1} &= JS^2a^2 \bm q^2 - \frac{J\hbar^2}{D^2} \omega^2
+\Sigma_1-\imu\Sigma_2 \hbar \omega
,
\end{align}
where the self-energy 
coefficients are given by
\begin{align}
\Sigma_1 &= \left( \frac{V_{\rm imp}^2S a^3}{2\pi J^{3/2} \ell^3} \right)^2
,\ \ \Sigma_2 = \sqrt{\frac{4J\Sigma_1}{D^2}}
, \label{eq:Sigma_3d_hm}
\\
\Sigma_1^{\rm (2D)} &= \frac{V_{\rm imp}^2S^2 a^2}{\pi J \ell^2}
,\ \ \Sigma_2^{\rm (2D)} = \sqrt{\frac{2J\Sigma_1}{D^2}}
\label{eq:Sigma_2d_hm}
\end{align}
The pinning length is given by $\xi_{\rm p} = a\sqrt{JS^2/\Sigma_1}$, which has the same form as SkX.
The coefficient is however different and the ratios between SkX and HMS cases are given by $\xi_{\rm p}^{\rm SkX}/\xi_{\rm p}^{\rm HMS} = \frac{2}{3}$ for $d=3$ and $\xi_{\rm p}^{\rm SkX}/\xi_{\rm p}^{\rm HMS} = \frac{\sqrt 2}{\sqrt 3}$ for $d=2$.
The shorter pinning length for SkX is due to the fact that the triple helices in SkX generate more pinning energy than the single helix state considered here.
Note that this value is dependent on the specific form of the pinning potential.

The equation of motion has the form of a damped oscillator, and the pinning frequency $\omega_{\rm p}$ and damping ratio $\zeta$ is determined as
\begin{align}
\hbar \omega_{\rm p} = D\sqrt{\frac{\Sigma_1}{J}}
, \ \ 
\zeta = \frac{ D\Sigma_2}{2\sqrt{J\Sigma_1}}
\end{align}
for three and two dimensions.
The pinning frequencies for HMS are listed in Tab.~\ref{tab:expr2}.
The damping ratio is estimated as $\zeta=1$ for $d=3$ and $\zeta=1/\sqrt{2}$ for $d=2$.
Namely, the HMS in $d=3$ is located at the critically damped case.

From the pole of the Green function, we can also have the dispersion relation 
\begin{align}
\hbar \omega (\bm q) &= DSa|\bm q|
\end{align}
for phason in the clean HMS.
With the disorder effects it becomes
\begin{align}
\hbar \omega' (\bm q) &\simeq \hbar \omega_{\rm p} 
+ \frac{DS^2a^2}{2} \sqrt{\frac{J}{\Sigma_1}} \, \bm q^2 + O(q^4)
.
\end{align}
We note that this excitation mode is accompanied by an attenuation.
Thus the nonreciprocity does not appear as distinct from the dispersion relations for SkX.

\begin{figure}[t]
\begin{center}
\includegraphics[width=50mm]{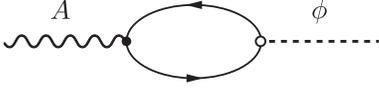}
\caption{
Illustration for the lowest-order coupling between phason field and vector potential.
}
\label{fig:diagram}
\end{center}
\end{figure}

\section{External field-spin coupling in single-helix case} \label{sec:appendix2}

We consider the phason-(uniform) external field coupling following the theory for CDW in one dimension \cite{lee74}.
We begin with the electron Hamiltonian
\begin{align}
\mathscr H_e &= \sum_{\sg\sg'}\int \diff \bm r
c^\dg_\sg(\bm r) 
\left\{ \left[\tfrac{1}{2m} {(-\imu\hbar \bm \nabla - e \bm A ) ^2} -\ep_{\rm F} \right] \delta_{\sg\sg'}
\right.
\nonumber \\
&\hspace{10mm} \left.
- J_{\mathrm H} \bm S(\bm r) \cdot \bm \sg_{\sg\sg'}
\right\} c_{\sg'}(\bm r)
\end{align}
where $\ep_{\rm F}=\frac{\hbar^2 k_{\rm F}^2}{2m}$ is the Fermi energy and $J_{\rm H}$ is the Hund coupling.
The coupling to $\eta$ ($\sim \omega\phi$) is here neglected, since it gives a higher-order contribution compared to $\phi$ field.
With this Hamiltonian, we can obtain the lowest-order contribution for the field-phason coupling which involves the electron-hole bubble shown in Fig.~\ref{fig:diagram}.
The bubble part $F_\mu$ connecting the vector potential $A_\mu$ and phason field $\phi$ is explicitly written in the Matsubara frequency domain as
\begin{align}
&F_{\mu} (\bm q,\imu\omega_m) = -4 e \Delta^2 k_{\rm B}T \sum_{n\bm k}
\nonumber \\
&\hspace{2mm} \times
\frac{(v^\mu_{\bm k}\bm v_{\bm k+\bm Q} - v^\mu_{\bm k+\bm Q}\bm v_{\bm k})\cdot \hbar \bm q
+ (v^\mu_{\bm k+\bm Q}-v^\mu_{\bm k})\imu \hbar \omega_m }
{[(\imu\hbar \ep_n-\ep_{\bm k})(\imu\hbar \ep_n-\ep_{\bm k+\bm Q})-4\Delta^2]^2}
\end{align}
where we have defined the fermionic Matsubara frequency $\hbar \ep_n=(2n+1)\pi k_{\rm B}T$, the mean-field $\Delta=\frac{1}{2}J_{\rm H}S$, the single-particle excitation energy $\ep_{\bm k}=\frac{\hbar^2\bm k^2}{2m} - \ep_{\rm F}$ and velocity $\bm v_{\bm k}=\frac{\hbar \bm k}{m}$.
We have kept only the leading order terms.
Assuming $\ep_{\rm F} \gg \Delta$ and $k_{\rm F} \gg Q$, 
the coupling term is given in the Hamiltonian form by
\begin{align}
\mathscr H_{E} = \frac{e\hbar^2 N''(0) \Delta^2}{3m a^d} \int \diff \bm r  \, \phi \, \bm Q \cdot \bm E
, \label{eq:e_field}
\end{align}
where we have introduced the electric field $\bm E = -\partial_t \bm A$ and the density of state $N(0)$ at the Fermi level.
We note the relation $N''(0) \sim 1/\ep_{\rm F}^3$.

The coupling between uniform magnetic field and spin is given by the Zeeman interaction \cite{tatara14}:
\begin{align}
\mathscr H_B &= -g_s\mu_{\rm B} \int\frac{\diff \bm r}{a^d} \bm B \cdot \bm S
\\
&= -\frac{g_s e\hbar S}{2mQa^d} \int \diff \bm r \, \eta \, \bm Q \cdot \bm B
.
\end{align}
Since the phason part has a fast spatially oscillating factor $\epn^{\imu Q y}$, the dominant contribution comes from the $\eta$-field.

\section{Dispersion relation for ferromagnetic state} \label{sec:appendix3}
Let us derive the dispersion relation for the excitation from the ferromagnetic state, to make a comparison with SkX and HMS.
The spin moment is given by $\bm S = S\hat {\bm z}$ in the ground state.
The classical spin-wave analysis gives the dispersion relation
\begin{align}
\hbar \omega(\bm q) &= g_s\mu_{\rm B}B + \frac{1}{2}JS a^2\bm q^2 
+ DS a q_z
\end{align}
which is obtained based on Eq.~\eqref{eq:ham_fm}.
In contrast to HMS, here the nonreciprocity along $z$ direction appears in the dispersion relation.

\end{document}